# Electrostatic Friction Force on an AFM probe Moving Near a Sample Surface

G. V. Dedkov[1] and A. A. Kanametov

Nanoscale Physics Group, Kabardino-Balkarian State University, Nalchik, 360004 Russia

**Abstract.** Within the framework of nonrelativistic electrodynamics, general formulas have been obtained for the tangential dissipative force of electrostatic friction and the normal force of attraction to the surface of an axially symmetric probe moving parallel to the smooth surface of homogeneous materials or coated with thin films substrates with various combinations of materials. As a numerical example, the interaction of a metal ball moving above a metal surface has been studied. The results of the calculations are compared with the available experimental and theoretical results of other authors.

**Introduction**

Electrostatic friction is one of the possible constituents of noncontact friction between two bodies in relative motion separated by a vacuum or an air gap. Static electric field between different surfaces may exist even without any externally applied voltage. The electrostatic attraction at zero bias voltage can be caused by local variations in the work function or surface contaminations resulting in an inhomogeneous electric field. This is referred to as the patch-charge effect [1–3]. Also, spatial fluctuations of the electric field can be caused by charged defects in the bulk of a probing body or a sample.

Electrostatic forces between conducting surfaces due to spatial variations in the surface potential were studied in [3]. The patch-potential variations were measured under various conditions using vibrating or rotating plate electrometers [4]. A lot of works were devoted to electrostatic forces in atomic force microscopy (AFM) [5–13] (the list is not complete). Measurements of electrostatic forces are important not only in AFMs, but also in precision measurements of the Casimir forces [14–16] and in designing MEMS devices [17–19].

Much less attention was paid to the problem of dissipative electrostatic forces, when a probing body is moving or oscillates near a surface [20–24]. Theoretical models of dissipative electrostatic forces were developed in [21, 24]. In these works, the tip of a metallic cantilever is considered as a section of a cylindrical surface with curvature radius *R*. The cantilever is perpendicular to a planar sample surface and the tip displacement is assumed to be parallel to the surface. Moreover, the authors also considered a model of a spherical [21, 24] or ellipsoidal [21] tip neglecting the spatial variations of the electric potential due to the curvature of the tip [24], or using the "proximity approximation" [21]. Along with that, simplified expressions for the tip-sample capacitance were used [24]. The friction coefficient was calculated via the Joule losses in the bulk of the sample [21], or through the Pointing vector on the surface of the sample [24]. Both methods led to close quantitative estimates of electrostatic friction and a similarity between the distance dependences and the tip radius. However, the numerical values of the friction forces turned out to be much less than in the experiment [20].

---

[1] Corresponding author e-mail: gv_dedkov@mail.ru

In this work, we carry out a direct calculation of the electrostatic force applied to a moving tip having a conical shape with a spherical extremity [5], or any other axially-symmetric shape. The distribution of charges on the tip is assumed to be fixed in the form of randomly distributed patch-charges or it is controlled by an external bias voltage $U$. The corresponding distribution of the induced charges on the sample is obtained by solving exact electrostatic problem with allowance for real shape of the tip and dielectric properties of the sample. The resultant friction force on the tip is then calculated similar to the calculation of the friction force on a point charge moving above a polarizable surface [25]. Since the probing tip in AFMs may have a very acute extremity, the charge density is maximal near its end and rapidly decreases with increasing distance to the sample. Correspondingly, an accurate determination of the charge distribution is crucially important in further calculations of the frictional force. We consider several combinations of the probe and sample materials, including metal substrates with thin dielectric films on the top and vice versa in order to reveal the effect of materials and to shed light on the role of thin films in electrostatic friction.

## 1. Normal and lateral electrostatic forces on a moving body with fixed distribution of charges

Figure 1 shows the sketch of the system and the used Cartesian coordinate system. A probing tip with fixed charges $q_j$ having instantaneous coordinates $(x_j + Vt, y_j, z_j)$ moves with the constant velocity $V$ above a sample with the dielectric function $\varepsilon(\omega)$. In a more general case (shown in the insert), the sample includes a thick substrate and a thin film with the thickness $d$ characterized by the dielectric functions $\varepsilon_2(\omega)$ (film) and $\varepsilon_3(\omega)$ (substrate). Within the nonrelativistic approximation $V/c \ll 1$, the electrodynamic problem reduces to solving the Poisson equation for the electrostatic potential $\phi(x, y, z, t)$:

$$\Delta\phi(x,y,z,t) = -4\pi\rho(x,y,z,t) = -4\pi\sum_j q_j \delta(x-(x_j+Vt))\delta(y-y_j)\delta(z-z_j) \quad (1)$$

Equation (1) should be solved under the conditions of continuity of potential $\phi$ and normal projection of the electric induction $-\varepsilon\, d\phi/dz$ at $z=0$ in the case of homogeneous sample, and at $z=0$, $z=-d$ in the case of a substrate and a film. Making use the Fourier-transforms of $\phi(\mathbf{r},z,t)$ and the charge density $\rho(\mathbf{r},z,t)$ in the right-hand side of (1) where $\mathbf{r}$ is a two-dimensional vector in the plane $(x,y)$, namely

$$\phi(\mathbf{r},z,t) = \frac{1}{(2\pi)^3}\int d^2k\, d\omega\, \phi_{\omega\mathbf{k}}(z)\exp[i(\mathbf{kr}-\omega t)] \quad (2)$$

$$\rho(\mathbf{r},z,t) = \frac{1}{(2\pi)^3}\int d^2k\, d\omega\, \rho_{\omega\mathbf{k}}(z)\exp[i(\mathbf{kr}-\omega t)] \quad (3)$$

with $\mathbf{k}=(k_x,k_y)$, one obtains from (1)

$$\left(\frac{d^2}{dz^2}-k^2\right)\phi_{\omega\mathbf{k}}(z) = -8\pi^2\,\delta(\omega-k_x V)\sum_j q_j \exp(-i\mathbf{k}\mathbf{r}_j) \quad (4)$$

Solving Eq. (4) with the corresponding boundary conditions (at $z=0$ and $z=-d$) yields

$$\phi_{\omega\mathbf{k}}(z) = \phi_{\omega\mathbf{k}}^{bare}(z) + \phi_{\omega\mathbf{k}}^{ind}(z) \quad (5)$$

$$\phi_{\omega\mathbf{k}}^{bare}(z) = -\frac{4\pi^2}{k}\delta(\omega-k_x V)\sum_j q_j \exp(-\mathbf{k}\mathbf{r}_j)\exp\left(-k|z-z_j|\right) \quad (6)$$

$$\phi_{\omega\mathbf{k}}^{ind}(z) = -\frac{4\pi^2}{k}D(\omega,k)\delta(\omega-k_x V)\sum_j q_j \exp(-\mathbf{k}\mathbf{r}_j)\exp(-k(z+z_j)) \quad (7)$$

$$D(\omega,k) = \frac{\Delta_1(\omega)-\Delta_2(\omega)\exp(-2kd)}{1-\Delta_1(\omega)\Delta_2(\omega)\exp(-2kd)} \quad (8)$$

$$\Delta_1(\omega) = \frac{\varepsilon_2(\omega)-1}{\varepsilon_2(\omega)+1},\quad \Delta_2(\omega) = \frac{\varepsilon_2(\omega)-\varepsilon_3(\omega)}{\varepsilon_2(\omega)+\varepsilon_3(\omega)} \quad (9)$$

The $\phi_{\omega \mathbf{k}}^{bare}(z)$ and $\phi_{\omega \mathbf{k}}^{ind}(z)$ in (5)–(7) correspond to the bare charge potential and the induced potential of the sample. Substituting (6), (7) into (2) and integrating by $\omega$ we get

$$\phi^{bare}(\mathbf{r},z,t) = -\frac{1}{2\pi}\sum_j q_j \int \frac{d^2k}{k} \exp\bigl(-k|z-z_j|\bigr)\exp\bigl(i\mathbf{k}(\mathbf{r}-\mathbf{r}_j)-ik_xVt\bigr) \qquad (10)$$

$$\phi^{ind}(\mathbf{r},z,t) = -\frac{1}{2\pi}\sum_j q_j \iint \frac{d^2k}{k} D(k_xV,k)\exp\bigl(-k(z+z_j)\bigr)\exp\bigl(i\mathbf{k}(\mathbf{r}-\mathbf{r}_j)-ik_xVt\bigr) \qquad (11)$$

The induced electric field is obtained using (2), (11) with allowance for the equation $\mathbf{E}^{ind} = -\nabla \phi^{ind}$. Summing the electrostatic forces $q_j \mathbf{E}^{ind}(x = x_j + Vt, y = y_j, z = z_j, t)$ on all point charges of the probe one obtains the following expressions for the resultant tangential and normal forces $F_x, F_y$:

$$F_x = \frac{i}{2\pi}\sum_j \sum_m q_j q_m \int d^2k \frac{k_x}{k} D(k_xV,k)\exp(-i\mathbf{k}(\mathbf{r}_j-\mathbf{r}_m))\exp(-k(z_j+z_m)) \qquad (12)$$

$$F_z = \frac{-1}{2\pi}\sum_j \sum_m q_j q_m \int d^2k\, D(k_xV,k)\exp(-i\mathbf{k}(\mathbf{r}_j-\mathbf{r}_m))\exp(-k(z_j+z_m)) \qquad (13)$$

where $\mathbf{r}_m = (x_m, y_m)$, $\mathbf{r}_j = (x_j, y_j)$. Formulas (10)–(13) have the general meaning irrespectively of the shape of the probe. In particular, for a single moving charge, Eqs. (10)–(13) reduce to the well-known results [25].

As follows from (12), (13), the electrostatic forces depend to large extent on the properties of response function (8). Therefore, it is worthwhile to examine several important combinations of the probe and sample materials. The friction force $F_x$ is determined by the imaginary part of $D(\omega,k)$, while the attraction force $F_z$ – by the real part of $D(\omega,k)$. Typical velocities of the AFM probes are very small even in dynamic modes ($V \ll 1\, m/s$), and at a typical distance of 10 $nm$ between the tip and the surface the main contributions to the integrals in (10)–(13) stem from the wave vectors of order $10^8\, m^{-1}$. Correspondingly, the frequencies $\omega = k_xV$ are of order $10^8\, s^{-1}$. Under these conditions, the imaginary parts of $D(\omega,k)$ and $\Delta_{1,2}(\omega)$ in (6), (7) will be much less than the real parts for any typical contacting materials (metals, semiconductors, ionic dielectrics, water films or their combinations). Then, with an accuracy of order $\left(\Delta_{1,2}''/\Delta_{1,2}'\right)^2$, from (6) we obtain

$$\mathrm{Im}(D(k_x V, k)) \approx \frac{\Delta_1''\left(1 - a^2|\Delta_2|^2\right) + a\Delta_2''\left(|\Delta_1|^2 - 1\right)}{\left(1 + a^2|\Delta_1|^2|\Delta_2|^2 - 2a\Delta_1'\Delta_2'\right)}, \tag{14}$$

$$\mathrm{Re}(D(k_x V, k)) \approx \frac{(\Delta_1' - a\Delta_2')(1 - a\Delta_1'\Delta_2')}{\left(1 + a^2|\Delta_1|^2|\Delta_2|^2 - 2a\Delta_1'\Delta_2'\right)}, \tag{15}$$

where $a = \exp(-2kd)$, one-primed and double-primed quantities denote the proper real and imaginary parts of the functions, $\Delta_{1,2}$.

According to the accepted law-frequency approximation, the dielectric functions of metals and dielectrics can be written in the form

$$\varepsilon(\omega) = 1 + i\frac{4\pi\sigma}{\omega} \tag{16}$$

$$\varepsilon(\omega) = \varepsilon_\infty + \frac{(\varepsilon_0 - \varepsilon_\infty)\omega_T^2}{\omega_T^2 - \omega^2 - i\gamma\omega} \approx \varepsilon_\infty + i\frac{(\varepsilon_0 - \varepsilon_\infty)\gamma\omega}{\omega_T^2} \tag{17}$$

where $\sigma$ is the metal conductivity, $\varepsilon_0, \varepsilon_\infty$ are the static and optical dielectric permittivity of a dielectric, $\gamma$ and $\omega_T$ – the damping factor and phonon frequency. The expressions analogous to (13), (14) can be also used for dielectrics with rotational relaxations. For example, for water one can use the fitting function [26]

$$\varepsilon(\omega) = \alpha + \frac{\beta}{1 - i\omega/\omega_0} \tag{18}$$

with $\alpha = 4.35$, $\beta = 72.24$, $\omega_0 = 1.3 \cdot 10^{11} \, s^{-1}$.

Using Eqs. (7), (11)–(15), it is easy to write the expressions for $\mathrm{Im}(D(k_x V, k))$ and $\mathrm{Re}(D(k_x V, k))$ for clean surfaces with dielectric functions (13)–(15) or various film-substrate combinations. These results are given in the Appendix. The corresponding formulas are applicable under the condition $\omega_0 z/V \gg 1$, where $\omega_0$ is the characteristic frequency ($2\pi\sigma$, $\omega_T$ or $\omega_0$ in the case of functions (13), (14)), and $z$ is the characteristic probe-sample distance.

## 3. Charge distribution and forces in the presence of bias voltage between the probe and the sample

Let the conductive probe is loaded by the external voltage $U$ and moves parallel to the grounded substrate: a homogeneous metal plate, a metal film on top of a dielectric plate, or a dielectric film on top of a metal plate. The general solution to the Poisson equation will be given by a sum of (10) and (11) provided that $\phi_S = U$ on the surface of the tip. It is this condition allows one to calculate the unknown charges $q_j$ distributed on the surface of the probe. When substituting the instantaneous coordinates $(\mathbf{r}, z) = (x + Vt, y, z)$ of any point on the surface of the probe into (10) and (11), the resultant potential $\phi_S$ is independent of time. So does the charge distribution. However, the induced charge of a sample varies in time and space.

In the practically important case of an axially symmetric probe (AFM tip, for example, Fig. 2), the surface charge distribution on the tip is also axially symmetric. Using this fact, we replace the discrete charges $q_m, q_j$ in (8)–(13) by the circular charges $q_m, q_j$ and divide them onto the discrete charges $q_{mn}$ and $q_{jp}$ corresponding to the rings $m$ and $j$. Taking advantage of the axial symmetry, we replace the additional summation by $n$ and $p$ by the angular integration. As a result, making use of the analytical properties of $D(\omega, k)$ (see Appendix), for the points $(r, z) \in S$ on the surface of the probe, Eqs. (10), (11) and Eqs. (12), (13) take the form:

$$\phi^{bare}(r,z) = \sum_j q_j \int_0^\infty dk J_0(kr) J_0(kr_j) \exp(-k|z-z_j|) = \sum_j q_j T(r, r_j, |z-z_j|) \qquad (19)$$

$$T(x, y, z) = \frac{2}{\pi} \frac{1}{\left[z^2 + (x+y)^2\right]^{1/2}} K\left[\left(\frac{4xy}{\left[z^2 + (x+y)^2\right]}\right)^{1/2}\right] \qquad (20)$$

$$\phi^{ind}(r,z) = -\frac{1}{2\pi} \sum_j q_j \int \frac{d^2 k}{k} D(k_x V, k) J_0(kr) J_0(kr_j) \exp(-k(z+z_j)) \qquad (21)$$

$$F_x = -\frac{1}{2\pi} \sum_m \sum_j q_m q_j \int d^2 k \frac{k_x}{k} \mathrm{Im}(D(k_x V, k)) J_0(kr_m) J_0(kr_j) \exp(-k(z_m + z_j)) \qquad (22)$$

$$F_z = -\frac{1}{2\pi}\sum_m\sum_j q_m q_j \int d^2k \, \text{Re}(D(k_x V, k)) J_0(kr_m) J_0(kr_j) \exp(-k(z_m + z_j)) \tag{23}$$

In (19)–(23), $K(x)$ is the elliptic integral of the first kind and $J_0(x)$ is the Bessel function.

By expanding $D(k_x V, k)$ (see Appendix) in powers of $(k_x V)$ and substituting into (21), it is easy to see that with an accuracy up to terms proportional to the velocity squared, the induced potential on the surface of the probe takes the form

$$\phi^{ind}(r,z) \approx -\sum_j q_j \int_0^\infty dk \, \text{Re}(D(0,k)) J_0(kr) J_0(kr_j) \exp(-k(z + z_j)) \tag{24}$$

Thus, in the small-velocity approximation, the induced potential of the probe coincides with the static induced potential corresponding to the static electrostatic problem for an immovable probe. The attraction force in this case is also independent on the velocity and matches the static values. Unlike this, the friction force depends on the time lag between the electric field of a moving charge and the sample response. This leads to the velocity-proportional dependence of $F_x$.

According to the results given in the Appendix,

$$\text{Re}(D(0,k)) = A\frac{p + \exp(-2kd)}{1 + p\exp(-2kd)} = A\sum_{s=1}^\infty (-1)^{s-1} p^{s-1} [p\exp(-2kd(s-1)) + \exp(-2kdn)] \tag{25}$$

where $p$ – parameter (see (A4)–(A6)). For clean surfaces $d \to \infty$ and $\text{Re}(D(0,k)) = p = const$. Substituting (25) into (24) yields

$$\phi^{ind}(r,z) = -p\sum_j q_j \sum_{s=1}^\infty (-p)^{s-1}[pT(r,r_j,2d(s-1) + z + z_j) + T(r,r_j,2ds + z + z_j)] \tag{26}$$

Following [12], to find the surface charges on the probe we use the least squares method and the condition $\sum_n (\phi_S - U)^2 = \min, 1 \leq n \leq N$, with $\phi_S = \phi^{bare}(r_n, z_n) + \phi^{ind}(r_n, z_n)$. The discrete points $(r(z_n), z_n)$ on the surface $S$ must be chosen according to the equation $r(z)$ of the surface. We introduce the grid $z_m$ on the $z$ axis for the circular charge distributions so that the points $z_n$ are located between the nodes of the grid $z_m$ ($1 \leq m \leq M$). The resulting system of linear equations for ring charges is solved by standard methods.

Knowing the ring charges $q_j$, and writing $D(k_x V, k)$ in the form

$$D(k_x V, k) \cong f_1(k) + i \cdot k_x V f_2(k)$$

where the functions $f_{1,2}(k)$ depend on material properties of the sample, (see Appendix), formulas (22) and (23) can be rewritten in the form being more convenient for numerical calculations

$$F_x = -\frac{V}{2} \sum_m \sum_j q_m q_j \int_0^\infty dk k^2 f_2(k) J_0(kr_m) J_0(kr_j) \exp(-k(z_m + z_j)) \qquad (27)$$

$$F_z = -\sum_m \sum_j q_m q_j \int dk\, k\, f_1(k) J_0(kr_m) J_0(kr_j) \exp(-k(z_m + z_j)) \qquad (28)$$

## 4. Numerical results

As an important numerical example, let us consider the calculations of electrostatic forces $F_x, F_z$ on a spherical metal nanoparticle (ball) with radius $R$ moving parallel to the surface of a metal plate at a speed $V$. The potentials of the surface of the probe and the surface are set equal to $U$ and zero, respectively. For the dielectric permittivity of both materials we use formula (16). In this case, in (27) and (28), $f_1 = 1$, $f_2 = 1/2\pi\sigma$. The total number of the circular charges was assumed to be $N = 300$. The grid spacing along the vertical coordinate was increasing exponentially in such a way that the ratio of the maximum and minimum step values was close to 20.

For further simplification, formulas (27) and (28) were reduced to a simpler form after introducing dimensionless charges $\tilde{q}_m = q_m / UR$, $\tilde{q}_j = q_j / UR$, and a new variable $u = z/R$:

$$F_x = -\frac{VU^2}{2R} f_0 \sum_{m,j} \tilde{q}_m \tilde{q}_j T_2(r_m/R, r_j/R, (z_m+z_j)/R) \qquad (29)$$

$$F_z = U^2 \sum_{m,j} \tilde{q}_m \tilde{q}_j T_1(r_m/R, r_j/R, (z_m+z_j)/R) \,, \qquad (30)$$

where the auxiliary functions $T_{1,2}(x, y, z)$ are defined by the relations

$$T_1(x,y,z) = \int_0^\infty dk k J_0(kx) J_0(ky) \exp(-kz) = \frac{d}{dz} T(x,y,z) \tag{31}$$

$$T_2(x,y,z) \equiv \int_0^\infty dk k^2 J_0(kx) J_0(ky) \exp(-kz) = \frac{d^2}{dz^2} T(x,y,z) \tag{32}$$

The function $K(x)$ in (20) was approximated in the form [27]

$$K(x) = \int_0^{\pi/2} \frac{dy}{(1-x^2 \sin^2 y)^{1/2}} \approx \sum_{n=0}^{4} \left( a_n \eta^n - b_n \eta^n \ln \eta \right), \quad \eta = 1 - x^2 \tag{33}$$

The polynomial fit in (33) has an error less than $2 \cdot 10^{-8}$. The coefficients $a_n, b_n$ are given in the Table.

Table

Fitting coefficients $a_n, b_n$ in (33), taken from [27]

| $n$ | $a_n$ | $b_n$ |
|---|---|---|
| 0 | 1.38629436112 | 0.5 |
| 1 | 0.09666344259 | 0.12498593597 |
| 2 | 0.03590092383 | 0.06880248576 |
| 3 | 0.03742563713 | 0.03328355346 |
| 4 | 0.01451196212 | 0.00441787012 |

Figure 3 shows the calculated values of the ring charges on the ball in relative units $q_n/RU$, depending on the number $n$ of rings. Curves 1, 2, 3 correspond to various relative distances of the ball apex from the surface: $z_0/R = 0.01, 0.05, 0.1$. The apparent increase in the charge density near the points most distant from the surface of the ball (at $n \approx 300$) is explained by the increase in the area of the rings because of the mesh unevenness.

The largest changes in the charge distribution occur near the apex of the ball facing the surface, while in the points of the ball that are most remote from the surface the changes in the charge distribution depend little on the minimum distance $z_0$ to the surface. Figure 4 shows the calculated forces $F_x$ and $F_z$ in relative units $VU^2/4\pi\sigma R$ and $U^2$ depending on the relative apex distance $z_0/R$. In this case, the unit value of $\tilde{F}_x$ corresponds to the absolute friction force of $0.478 \cdot 10^{-10} \, nN$ (at $V = 1 \, m/s, U = 1 V, R = 10 \, nm, \sigma = 1.85 \cdot 10^{17} \, s^{-1}$ for Au), while the unit value of $\tilde{F}_z$ corresponds to the absolute value of $1/9 \, nN$. We note that the values of $V, U$ and $R$ in (29), (30) must be taken in Gaussian units. Using the above dependences (curves 1 and 2), it is easy to find the values of the forces for any distances and other parameters of the problem.

It is of interest to compare the results of calculating the frictional force with experiment [20], in which the radius of the spherical tip of the probe (Au) was equal to $1 \, \mu m$ (for the same values of the remaining parameters). At $z_0 = 20 \, nm$ and $T = 300 K$, the measured dissipative force proportional to the velocity of the probe was $F_x \sim 3 \cdot 10^{-12} \, N$. In our case, using the graph in Fig. 4, we obtain the value $F_x \sim 1.8 \cdot 10^{-20} \, N$, and the absolute dependence on the distance is close to $F_x \sim z_0^{-1.5}$ at $10 \leq z_0 \leq 30 \, nm$ (as in the experiment [20]). At smaller distances from the surface, the exponent $n$ in the dependence $F_x \sim z_0^{-n}$ is closer to 2. The absolute value of the friction force is about three times higher than follows from the calculation results in [24] (for the same values of other parameters), but also significantly less than the value of the dissipative force measured in the experiment [20]. Thus, the problem of interpreting dissipative forces in [20] remains open.

## 5. Conclusions

In the framework of nonrelativistic problem of electrodynamics, general formulas are obtained for the tangential dissipative force of electrostatic friction and normal force of attraction to the surface of an axially symmetric probe moving parallel to the surface of homogeneous materials or coated with thin films of substrates with various combinations of materials. It is shown that the results of calculating the frictional force of a spherical probe for the conductive materials of the probe and the surface can be represented in a universal form. Comparison of the numerical results with the available experimental values of the dissipative forces under the conditions of electrostatic interaction reveals a smaller value with a discrepancy of 8 orders of magnitude, as in the theoretical calculations of other authors. In our opinion, taking into account the nature of the theoretical dependence of $F_x$ on the material parameters, it is more expedient to use the samples (plates) with low conductivity (homogeneous or with dielectric coatings) for the experimental study of electrostatic friction.

**Appendix**

Consider several possible combinations of materials, using Eqs. (7), (10)–(14). The corresponding functions $\text{Re}(D(\omega,k))$ and $\text{Im}(D(\omega,k))$ are given by:

*Clean surfaces*

a) metal

$$\text{Re}\,D = 1, \quad \text{Im}\,D = \frac{\omega}{2\pi\sigma} \tag{A1}$$

b) dielectric (13)

$$\text{Re}\,D = \frac{\varepsilon_\infty - 1}{\varepsilon_\infty + 1}, \quad \text{Im}\,D = \frac{2(\varepsilon_0 - \varepsilon_\infty)\gamma\omega}{(\varepsilon_\infty + 1)^2 \omega_T^2} \tag{A2}$$

c) dielectric (14)

$$\text{Re}\,D = \frac{\alpha + \beta - 1}{\alpha + \beta + 1}, \quad \text{Im}\,D = \frac{2\beta}{(\alpha + \beta + 1)^2}\frac{\omega}{\omega_0} \tag{A3}$$

*Thin films of thickness $d$ on a thick substrate*

a) metal (top) – dielectric (13) or (14) (bottom)

$$\text{Re}\,D = 1, \quad \text{Im}\,D = \frac{1 - \exp(-2kd)}{1 + \exp(-2kd)}\frac{\omega}{2\pi\sigma} \tag{A4}$$

b) dielectric (13) (top) – metal (bottom)

$$\text{Re}\,D = \frac{\exp(-2kd) + \dfrac{\varepsilon_\infty - 1}{\varepsilon_\infty + 1}}{1 + \exp(-2kd)\dfrac{\varepsilon_\infty - 1}{\varepsilon_\infty + 1}}, \quad \text{Im}\,D = \frac{(1 - \exp(-4kd))}{\left(1 + \exp(-2kd)\dfrac{\varepsilon_\infty - 1}{\varepsilon_\infty + 1}\right)^2}\frac{2\gamma\omega}{\omega_T^2} \tag{A5}$$

c) dielectric (14) (top) – metal (bottom)

$$\text{Re}\,D = \frac{\left(\dfrac{\alpha + \beta - 1}{\alpha + \beta + 1} + \exp(-2kd)\right)}{\left(1 + \exp(-2kd)\dfrac{\alpha + \beta - 1}{\alpha + \beta + 1}\right)}, \quad \text{Im}\,D = \frac{(1 - \exp(-2kd))}{(1 + \exp(-2kd))}\frac{2\beta\omega}{(\alpha + \beta + 1)^2} \tag{A6}$$

Fig. 1. Geometric configuration and coordinate system used in the calculation.

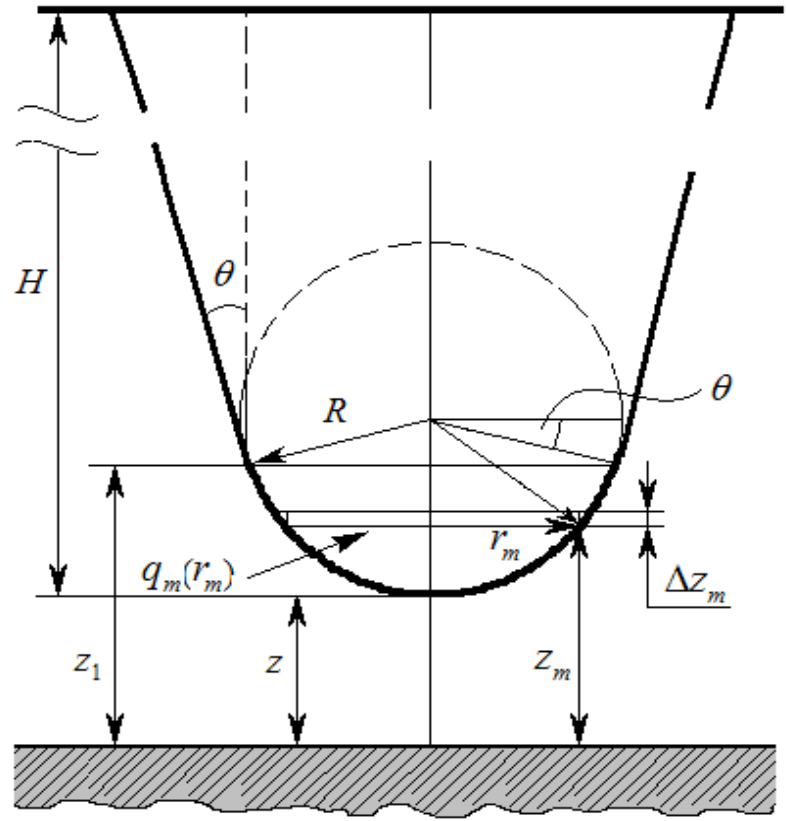

Fig. 2. Schematic view of the conical probe with a spherical extremity characterized by the parameters $R$, $H$, $\theta$; $r_m$ – radius of the m-th circular ring on the surface with charge $q_m$, $\Delta z_m$ – the step of mesh corresponding to the m-th ring, $z_1$ – the height of the point of the probe above the surface where the spherical extremity is conjugated with conical surface

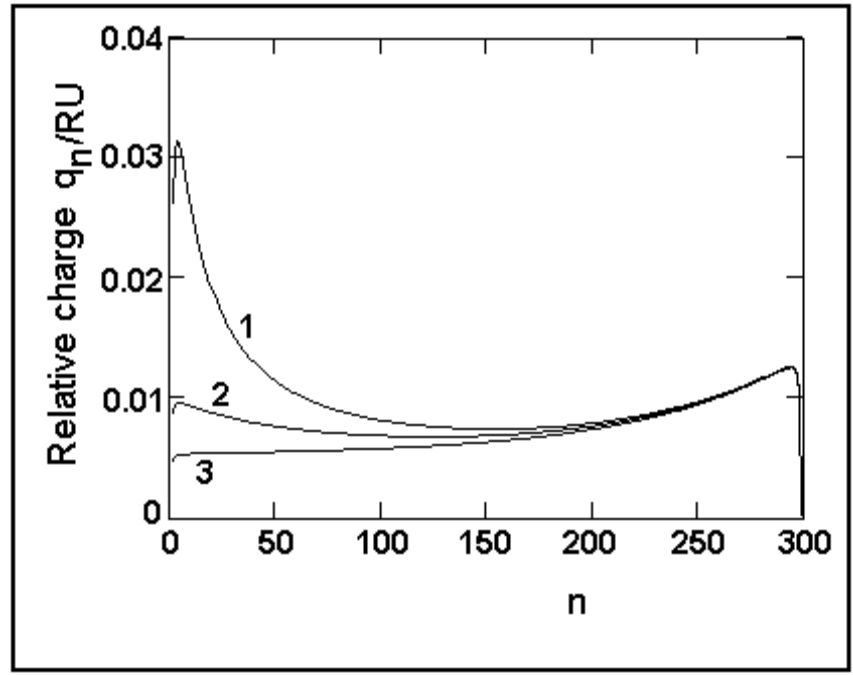

Fig. 3. Distribution of ring charges on the surface of a metal ball. Lines 1-3 correspond to the distance of the apex of the ball from the surface of 0.01, 0.05 and 0.1 (in units of radius). The frequency of ring distributions progressively decreases with growth approximately 20 times as one moves from bottom to top.

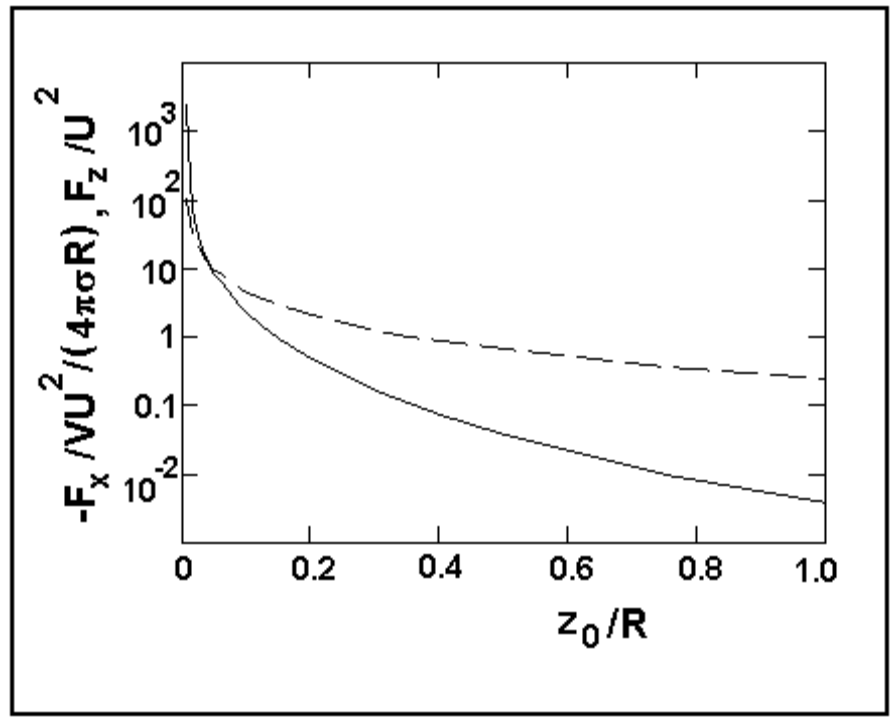

Fig. 4. Forces $F_x$ (solid line) and $F_z$ (dashed line), expressed in relative units, depending on the ball apex distance from the planar surface. To obtain the absolute values of the forces (in dynes), the parameters $V, U, R$ and $\sigma$ must be expressed in Gaussian units.